# *Identification of PAH Isomeric Structure in Cosmic Dust Analogues: the AROMA setup*

Hassan Sabbah[1,2,3], Anthony Bonnamy[1,2], Dimitris Papanastasiou[4], Jose Cernicharo[5], Jose-Angel Martín-Gago[5] and Christine Joblin[1,2, *]

[1] Université de Toulouse, UPS-OMP, Institut de Recherche en Astrophysique et Planétologie, 9 avenue du Colonel Roche, 31028 Toulouse Cedex 4 (France)
[2] CNRS, IRAP, 9 avenue du Colonel Roche, 31028 Toulouse Cedex 4 (France)
[3] CNRS, LCAR, IRSAMC, 118 route de Narbonne, 31013 Toulouse Cedex 6, (France)
[4] Fasmatech Science + Technology, Athens, GR
[5] Instituto de Ciencia de Materiales de Madrid, Sor Juana Inés de la Cruz, 3, Cantoblanco, 28049 Madrid, Spain

*Corresponding author: christine.joblin@irap.omp.eu

## Abstract

We developed a new analytical experimental setup called AROMA (Astrochemistry Research of Organics with Molecular Analyzer) that combines laser desorption/ionization techniques with ion trap mass spectrometry. We report here on the ability of the apparatus to detect aromatic species in complex materials of astrophysical interests and characterize their structures. A limit of detection of 100 femto-grams has been achieved using pure polycyclic aromatic hydrocarbon (PAH) samples, which corresponds to $2 \times 10^8$ molecules in the case of coronene ($C_{24}H_{12}$). We detected the PAH distribution in the Murchison meteorite, which is made of a complex mixture of extraterrestrial organic compounds. In addition, collision induced dissociation experiments were performed on selected species detected in Murchison, which led to the first firm identification of pyrene and its methylated derivatives in this sample.

# 1. Introduction

Organic material is ubiquitous throughout the Galaxy and Solar system. To date, around 200 molecules (see, e.g., http://www.astro.uni-koeln.de/cdms/molecules), often organic in nature, have been detected towards different lines of sight of varying physical and chemical conditions leading to interest in their origin and synthesis (Herbst and van Dishoeck, 2009). In addition to these molecular species, cosmic dust is also observed in a wide variety of astrophysical environments that include circumstellar envelopes, dense interstellar clouds, and protoplanetary disks. Dust is also found in less molecular environments such as the diffuse interstellar medium and HII regions. These are evidence of the rich and complex chemistry occurring in interstellar and circumstellar environments. While cosmic molecules are formed under a large range of physical conditions, refractory dust grains can only be efficiently formed in the hot regions of evolved stars, mainly Asymptotic Giant Branch (AGB) stars and more massive stars producing supernovae (SNe), which constitute the only possible sites for dust formation in the early Universe.

Spectral signatures of these grains that have been produced in circumstellar environments of dying stars are now found in Solar System objects (Kwok, 2004). Moreover, the discovery of pre-solar grains in meteorites and interplanetary dust particles (IDPs) made by observing isotopic anomalies, suggests that the organic material in these objects has an interstellar or proto-stellar heritage (Keller *et al.*, 2004).

The two major chemical families of cosmic dust components are the oxides (largely silicates) and the carbonaceous grains of which polycyclic aromatic hydrocarbons (PAHs) are the smallest in size and could be the building block of the larger carbonaceous cosmic materials. Their characteristic infrared emission bands at 3.3, 6.2, 7.7, 8.6, and 11.2 μm, the so-called Aromatic Infrared Bands (AIBs), dominate the mid-infrared spectrum of regions of planet and star formation (Léger & Puget, 1984; Allamandola & Tielens, 1985; Tielens, 2008). The proposed mechanism is that these AIBs arise from vibrational cooling cascades following heating of PAHs by the absorption of UV photons. PAHs are found to be the most abundant complex polyatomic molecules in the interstellar medium, accounting perhaps for as much as 10-20% of all carbon in our galaxy (Joblin et al., 1992; Draine, 2003) . The shape and intensity of the AIB spectra present some variations from one region to another, probably reflecting a variety of molecular sizes, structures and abundances (Tielens, 2008). At the present time the exact molecular identification of these aromatic species remains uncertain.

Most of our knowledge on the chemical composition and evolution of carbonaceous cosmic materials is based on astronomical observations. In addition, the analysis in the laboratory of organic extraterrestrial matter from meteorites, IDPs and materials from sample return missions provides additional clues on the composition and origin of these grains. Learning more about their formation, destruction, and alteration processes in astrophysical environments has motivated a number of research groups to simulate these processes in the laboratory. A large variety of reactors have been employed to produce cosmic dust analogues among which premixed (Buchta *et al.*, 1995), diffusion (Pino *et al.*, 2008), and atmospheric flames (Mennella et al., 1995), laser induced pyrolysis of hydrocarbon mixtures ( Schnaiter et al., 1999; Galvez et al.,  2002; Jäger et al., 2006;), plasma discharge of hydrocarbons ( Sakata et al., 1984; Contreras & Salama, 2013), and laser ablation of graphite targets (Scott and Duley, 1996) as well as pyrolysis of acetylene in a porous graphite reactor (Biennier *et al.*, 2009). In this context, a new experimental setup called the Stardust machine is being developed, at the CSIC, Madrid, to produce cosmic dust analogues from a wide variety of metals and organic precursors under well controlled physical conditions, approaching those encountered in evolved star environments.

Meteorites are the most available type of extraterrestrial material on Earth and are representative of the early Solar System. Much of our current understanding of meteoritic organic matter has come from investigations on the Murchison carbonaceous chondrite ( Sephton, 2002; Sephton, 2005). Analysis of this meteorite has shown that the organic content can be classified into two categories. The first and major fraction of organics is an insoluble macromolecular material of complex composition present as kerogen-like macromolecules, a highly cross-linked aromatic network (Pizzarello and Shock, 2010). The soluble organic portion is made up of tens of thousands of different molecular organic species (Schmitt-Kopplin *et al.*, 2010). Compound classes include aliphatic hydrocarbons, aromatic hydrocarbons, amino acids, carboxylic acids, sulfonic acids, phosphonic acids, alcohols, aldehydes, ketones, sugars, amines, and amides (Sephton, 2005). This motivates studies on the potential of this type of material for prebiotic chemistry (Pizzarello and Shock, 2010). From this complexity, PAHs predominate in abundances and are often present with their alkylated derivatives which could result from aqueous alteration in the parent body (Martins *et al.*, 2015).

Significant advances in the understanding of the meteoritic organic composition are tightly associated with improvements in analytical techniques. Cosmic dust grains are µm-sized or less. For this reason, the suitable analytical technique for their study should be sensitive and allow for high spatial resolution. Two-step laser desorption laser ionization mass spectrometry (L2MS) has been shown to be a valuable technique for selectively probing the molecular aromatic composition of solid materials, both of terrestrial and extraterrestrial origins (Hahn et al., 1987; Kovalenko et al., 1992; Spencer et al., 2008; Faccinetto et al., 2008; Sabbah et al., 2012). The L2MS technique requires minimal sample processing and handling, thus minimizing the risk of additional sample contamination. It has been previously used to assess terrestrial contamination of both meteorites (Plows et al., 2003) and Stardust Mission aerogel capture media (Sandford *et al.*, 2006; Spencer and Zare, 2007). Additionally, owing to the low detection limits of this technique ($10^{-18}$ mole regime) (Hahn, Zenobi and Zare, 1987) the sample size can be very small and the analysis leaves the bulk of the sample unaltered and available for further characterization. As an example of such studies, PAH compounds were unexpectedly detected within the Almahata Sitta meteorites and their distribution within meteorite fragments provided important clues as to the identity and history of the asteroid parent body (Sabbah et al., 2010). Moreover, one can achieve micrometer spatial resolution with the L2MS so that the organic measurements can be correlated with mineralogical features of the studied sample.

Most of the existing L2MS experimental setups employ coaxial time of flight (TOF) mass spectrometry. Orthogonal TOF (oTOF) decouples the conditions established in the laser desorption/ionization source with the initial conditions for TOF mass analysis, so enhanced mass resolving power can be obtained over the entire mass range of interest. While a TOF is a very powerful mass spectrometry tool and has been successfully combined with L2MS in the past (Spencer, Hammond and Zare, 2008; Faccinetto *et al.*, 2011), it does not provide structural information on the studied species. This is achievable via a two-step (MS/MS) experiments in which ions are first stored in a trap and then are fragmented under the action of photon or collision activation. The resulting fragments are then detected by mass spectrometry providing information on the molecular structure of the parent species.

In this work, we present a new experimental setup called AROMA (Astrochemistry Research of Organics with Molecular Analyzer) which has been developed to characterize the molecular content of cosmic dust analogues. We then describe the performances of this instrument for the detection of PAHs in pure samples and in complex samples such as the Murchison meteorite. The identification of pyrene ($C_{16}H_{10}$) and its methylated derivatives by MS/MS experiments in Murchison is demonstrated and the results are discussed.

## 2. The AROMA setup

AROMA is a unique molecular analyzer setup developed in the framework of the Nanocosmos ERC synergy project. The aim of Nanocosmos is to understand the physical and chemical processes leading to the formation of cosmic dust. The AROMA main purpose is to analyze, with micro-scale resolution, the molecular content of cosmic dust analogues, including the stardust analogues that will be produced in the Nanocosmos Stardust machine in Madrid. TOF mass spectrometers have evolved rapidly in recent years and, particularly oTOF systems, are currently capable of providing high mass resolution (m/Δm = $10^4$). This is by far more than any existing L2MS systems with mass resolutions in the order of a few hundreds, typically limited by the initial conditions in the ionization source. Our experimental setup combines the features of ion trap mass spectrometry, especially the ability to perform MS/MS experiments, with the very high sensitivity and high mass resolution provided by oTOF-MS. Most importantly, AROMA allows for the characterization of the structure of the detected molecules, which can provide insights for the disentanglement of isomers and is unmatched by other existing similar experimental setups. This is done by achieving MS/MS experiments on detected species employing collision induced dissociation (CID) or UV photodissociation studies. The construction of the AROMA setup was performed by Fasmatech, Greece, following requirements from the IRAP scientific team.

The AROMA platform consists of the following principal components (see Figure 1): a microprobe laser desorption ionization chamber accommodating a set of DC high-voltage lenses for collimating ions, a radio-frequency (RF) octapole cooler for receiving and thermalizing laser produced ions, a segmented linear quadrupole ion trap (LQIT) for storing and processing ions, a set of RF-DC optics for transferring ion pulses to high vacuum and finally an oTOF mass analyzer equipped with a two-stage reflectron and a fast microchannel plates (MCP) detector. The ionization source offers the possibility of probing solid samples by performing Laser Desorption Ionization (LDI) in a single or a two-step scheme, the latter involving desorption and ionization which are separated in time and space and performed by two different lasers respectively. All the data presented in this work have been produced using two-step LDI, and therefore in the L2MS configuration.

Samples are positioned vertically in the ion source chamber at low pressure ($10^{-6}$ mbar) and are observed by a microscopic camera. Chemical species in solid phase are desorbed from the sample platter using the fundamental output of a Nd:YAG pulsed laser at 1064 nm (Q-Smart 450, Quantel). They form an expanding plume in the vacuum which lasts for few microseconds. This pulse has a 5 ns duration and the laser beam is focused to a spot of 150 μm diameter. The typical pulse energy used for this experiment is 40-500 μJ/pulse, resulting in a fluence of 0.02 J/cm$^2$. Pulsed laser light focused on a highly localized area results in a heating rate of about $10^8$ K/s, inducing a much more rapid temperature rise than resistive heating, usually in the range of 10-50 K/s (Deckert and George, 1987). This rapid heating process favors desorption over decomposition, allowing this technique to desorb many neutral species with minimal fragmentation. At the basis of the plume the density of molecules can be very large allowing three body reactions and modifying chemical composition of the desorbed molecules. Such regime has not been observed in our experimental tests on pure PAH compounds (see Figure 2). The plume is then intercepted perpendicularly by a second laser in the UV domain using the 4$^{th}$ harmonic output at 266 nm of a Nd:YAG laser. The typical pulse energy used for this experiment is 3 to 5 mJ/pulse. This leads to selectively ionize the aromatic species that are present in the plume that can undergo (1+1) resonance-enhanced multiphoton ionization (REMPI). Gas-phase ionization creates a low-density cloud of charged particles not commonly considered as a plasma.

Ions generated in the plume are guided by a set of high-voltage lenses and focused through a narrow orifice into a differentially pumped RF cooler combining an octapolar-field component at the entrance to enhance acceptance and and a quadrupolar-field component at the exit to improve transmission. The arrival time of the ions to the RF cooler is synchronized with a gas pulse, typically helium or argon, raising pressure to ~$10^{-2}$ mbar during a 5-10ms time window to kinetically thermalize the ions. A weak DC gradient applied across the three segments of the cooler is used to transfer ions through a narrow aperture into the segmented LQIT. Radial confinement of ions in the octapole cooler and the LQIT is accomplished by a pair of rectangular RF waveforms at 180º out-of-phase and set at 180$V_{0p}$ and variable frequency. The LQIT is designed with eight variable length segments (Q1-Q8) where each segment can be switched between eight different DC levels during the course of an experiment. Lossless transfer of ions between different regions of the LQIT is accomplished by appropriate selection of DC settings forming gradients and axial potential wells. Ions are originally injected and stored in the Q2 segment in which isolation and CID can be performed. Ion isolation is accomplished either by the application of a resolving DC applied to the poles of the Q2 segment or by the application of a filtered-notch-field (FNF). Slow heating CID is performed under the influence of a sinusoidal waveform applied in dipolar form and superimposed to the rectangular RF waveforms. The Q5 segment is designed with optical access to to allow for photodissociation experiments to be performed. Chemical reaction studies between stored ions and injected neutrals is also possible in both segments. At the end of a processing cycle ions are transferred at the end-segment Q8 and ejected with 25eV energy through a RF hexapole ion guide and a low-voltage DC lens to the oTOF extraction region.

## 3. **Results**

### *3.1 Performances on pure PAH samples*

The experiments reported here focus on demonstrating the capability of our setup to identify aromatic molecules in cosmic dust analogues. This includes the coupling of the two-step LDI technique with the segmented linear ion trap, the CID experiments performed on individual PAH molecules and on a complex meteoritic sample and the overall performances of the instrument in terms of sensitivity and mass resolving power. All chemical compounds used for calibration and sensitivity tests were purchased from Sigma-Aldrich (St. Louis, MO). The PAH compounds were dissolved to 1 mg/mL solutions in toluene and further diluted, if needed for the sensitivity tests. From each solution, a 5 μL drop was spotted onto a 6 mm stainless steel sample disc. These discs are then attached to the sample holder. All samples were introduced into the system via a vacuum interlock, after allowing 1/2 h for the toluene to evaporate under ambient conditions.

Successful measurements have been carried out for a series of aromatic molecules, including toluene ($C_7H_8$), naphthalene ($C_{10}H_8$), pyrene ($C_{16}H_{10}$), methyl-pyrene ($C_{17}H_{12}$), pentacene ($C_{22}H_{14}$), coronene ($C_{24}H_{12}$) and fullerenes $C_{60}$. Figures 2a and 2b show typical mass spectra of pyrene and coronene molecules, respectively, showing that the parent ion peak largely dominates in each spectrum. Figure 2b shows also the impurity in the coronene sample which is the benzo(ghi)perylene ($C_{22}H_{12}$) at m/z=276. Small intensity fragmentation peaks (less than 10% of the parent ion peak) corresponding to dehydrogenated series (-H and -2H) were observed. The lack of serious fragmentation is important for the identification of a PAH distribution contained in a complex sample. For mass spectra presented here the mass resolution (m/Δm) is 8000. The limits of detection for these compounds were measured to be in the order of 100 femto-grams per laser shot (100 attomoles). For this purpose 50 ng of each compound has been deposited on the 6mm disc using a diluted solution of 10μg/mL.

Separating the desorption and the ionization processes by using two different lasers allows us to optimize the detection of aromatic species by changing three parameters: the IR laser energy per pulse, the UV energy per pulse and the delay time between the two laser pulses. Figures 3a and 3b show the variation of the pyrene cation parent peak intensity as we vary one of the laser pulse energy while fixing the second one. Both graphs show a linear trend. Signal fluctuations from one IR energy to the other are likely related to the expansion dynamics of the plume, which results in the UV pulse probing a different concentration of neutral desorbed molecules. Figure 3c shows the variation of the pyrene and coronene parent peaks as a function of the delay between desorption and ionization steps. It shows that the optimized delay time for typical experiments is around 1.3 μs. Figure 3d shows two mass spectra of pyrene recorded at the minimum and maximum IR laser pulse energy used for these experiments. In the range of energy explored for the desorption and ionization lasers, we did not observe significant change due to fragmentation.

### *3.2 Extraterrestrial materials: Murchison*

For the present work a fragment of one milligram of Murchison was powdered in order to be analyzed. The powder was attached to the sample holder using a 6 mm disc of carbon conductive tabs, double coated, which does not contribute to the background signal. The typical laser pulse energies used for this experiment are 150 μJ/pulse for the IR and 4mJ/pulse for the UV. Before acquisition, the mass spectrometer is externally calibrated by considering the parent ions produced by the two-step LDI of a standard calibration mixture of toluene ($C_7H_8$), pyrene ($C_{16}H_{10}$) and coronene ($C_{24}H_{12}$). After the acquisition, the obtained mass spectrum was internally recalibrated using several unambiguously identified peaks. The first step in the characterization of organics in a natural sample is to determine the molecular mass distribution. Figure 4 shows the AROMA mass spectrum of the Murchison bulk analysis including some peak assignments (based on possible isomer structures). We found a distribution of aromatic species ranging from m/z =116 to 300 consisting of PAHs and some related products, spanning size from one to seven benzoic cycles. This distribution can be compared to the one earlier published by Callahan et al. (2008) using a similar two-step LDI scheme. Whereas similar species are observed by their m/z positions, the distribution peaks at m/z=202 in the present work compared to m/z=228 in the previous one. Also, we did not detect masses over m/z=300, whereas these authors reported minor species up to m/z=374. These discrepancies can be explained in terms of homogeneity of the Murchison meteorite, sample preparation and instrument response. Callahan et al. prepared sublimates from Murchison acid residue obtained at 200°C and 600°C. This process could lead to thermal alteration of the species especially at the highest temperatures used to reveal the largest masses. In our process, we use directly a powdered Murchison sample. Finally, relative peak intensities depend on several factors, including desorption and photoionization efficiencies as well as the instrument response. Desorption characteristics have been found in previous work to depend on the nature of the matrix of the sample analyzed. Moreover, the ionization efficiency of neutral desorbed molecules depends on ionization cross-sections. These reasons make it difficult to obtain absolute peak intensities, which could be related to molecular abundances within the solid materials. To conclude our spectrum is consistent with the ones obtained by Callahan et al. but with the advantage of a better mass resolution and S/N ratio, particularly for masses above 200.

The elemental composition of each detected signal with a signal-to-noise ratio (S/N) greater than 10 was determined. This is achieved employing mMass software (Strohalm *et al.*, 2010), an open source mass spectrometry tool. A list of detected species by their mass and exact molecular formula is given in Table 1. Peaks which are positioned at odd mass include the $^{13}C$- containing isotopic homologues of the molecules as well as fragments resulting from dehydrogenation. High dehydrogenation behavior has been observed for small aromatic species, for instance peaks at m/z = 116 and 140 have their [M-H]$^+$, respectively m/z at 115 and 139, much higher in intensity than the M$^+$. This behavior could be explained by the fact that the matrix of Murchison is black and efficiently absorbs the IR radiation of the desorption laser leading to highly excited desorbed molecules, which can lose hydrogen once ionized. Another possibility could be that the radical cations [M-H]$^+$ result from the interaction of aromatic species with metals present in the highly heterogeneous matrix of Murchison.

To obtain possible molecular structure assignments, we first identify the masses by their chemical formula corresponding to the most simple PAH skeletons. Some possible isomeric structures of these molecules are shown in Figure 4. We then calculate the double equivalent bond (DBE) for each molecule. The DBE is representative of the unsaturation level of the molecules and thus corresponds to a direct measure of their aromaticity. The DBE method is used in several fields, as in petroleomics to separate and sort petroleum components according to their heteroatom class ($N_nO_oS_s$) and carbon number (Marshall and Rodgers, 2008). The DBE value is equal to the number of rings plus double bonds involving carbon atoms (because each ring or double bond results in a loss of two hydrogen atoms):

$$\text{Double bond equivalents } (C_cH_hN_nO_oS_s) = c - h/2 + n/2 + 1. \quad [1]$$

A decreasing number of H atoms in a molecule increases unsaturation and hence leads to higher DBE values. DBE is independent of the number of oxygen and sulfur atoms in the molecules. DBE values for each detected mass are directly provided by mMass. They are listed in table 1. DBE versus C number plot is used in mass spectrometry to simplify the visualization and understanding of complex hydrocarbon mixture. In figure 5 we represent the DBE versus C number for all the detected species. Several series of compounds are aligned horizontally and spaced by one or n carbon numbers. A homologous series of compounds with different degrees of alkylation (substitution of peripheral H by $CH_3$ groups leading to no change in the DBE) has constant DBE value but different carbon numbers. This type of series is represented in figure 5 by dashed lines. We observe a series of condensed PAHs spanning size from two to seven aromatic cycles accompanied by a series of methylated aromatic species. PAHs with alkyl side groups show more fragmentation, by losing hydrogen atoms, than parent species. Peaks at m/z = 116, 140 and 150 have been assigned respectively to ethynyl-toluene, diethynyl-methyl-benzene and triethynyl-benzene. These three species have at least one aromatic cycle, based on their DBE values, and suffer from extensive dehydrogenation as seen by the group of peaks in Figure 4. They can be related to benzene, toluene, and alkyl-benzenes that have been detected in Murchison using a thermal extraction method coupled with gas chromatography mass spectrometry (Sephton, 2002).

We have illustrated how the DBE representation can be used to simplify the analysis of a complex mixture through classification. In the following section, we will show how we can determine the structure of the peak at m/z=202, which corresponds to the most intense peak in our spectrum and is related to four other compounds in the DBE plot.

### 3.3 Structural identification

In order to investigate the structures corresponding to m/z=202 ($C_{16}H_{10}^+$) we performed CID experiments on both the Murchison sample and a pure pyrene sample. The $C_{16}H_{10}^+$ species from the pyrene sample are produced in the low pressure source ($10^{-6}$ mbar) by the two-step LDI technique and thermalized in the RF octapole cooler before being stored in the LQIT. Once stored, parent ions with m/z=202 are isolated. CID is then performed in the LQIT, while pulsing

He gas (peak pressure around few $10^{-2}$ mbar for a 5 ms period) and applying a dipolar excitation (DE) voltage to excite ion motion. The efficiency of CID and production of sequential generations of fragments depend on the DE frequency and voltage; it increases close to resonance and/or by increasing excitation voltages. Dissociation products are monitored by the oTOF.

Figure 6 shows the CID spectra performed on isolated cations at m/z=202 from a pyrene sample and from Murchison powder. Some fragment peaks are assigned with formulas to show the losses of hydrogen and carbon. Peaks related to hydrogen loss dominate both spectra. This behavior has been observed previously by West et al. (2014), where we studied the photodissociation of pyrene cations. Peaks related to the loss of one and two $C_2H_2$ from the parent peak has been also detected in this previous study. Figure 6 shows that both samples show a very similar dissociation pattern, which confirms that pyrene is the main PAH present in Murchison. Slight deviations between the two spectra can be associated to the manipulation of the ions from production to CID. In particular, the peak at m/z = 193 is likely a by-product. We tentatively assigned it to $C_{14}H_9O$ which would be formed by reactivity of $C_{15}H_9$ (m/z = 189) with $O_2$ in the residual background gas.

## 4. <u>Conclusion and perspectives</u>

We succeeded in coupling a two-step LDI technique to a segmented LQIT-oTOF to analyze the aromatic molecular content of cosmic dust analogues. This is the first time that two-step LDI is coupled to a linear ion trap with MS/MS capabilities. We achieved a limit of detection of ≈100 femto-grams with a single laser shot applied on pure PAH samples and of pico-grams for complex natural samples. We detected the PAH distribution in the Murchison meteorite and the results are in very good agreement with previous works with optimized performances in sensitivity and mass resolution. The production technique of aromatic ions in AROMA is a direct way of analysis compared to thermal or solvent extraction. In addition, by performing CID experiments we firmly identified the main peak at m/z=202 as due to pyrene. Combining CID experiments and DBE plot representation, we identified a series of methylated pyrene species.

AROMA setup, being highly sensitive, selective, spatially resolved, and owing the MS/MS capabilities enables unique chemical characterization of aromatic species in cosmic dust analogues and extraterrestrial samples. Changing the ionization source will enlarge the scope of investigatd chemical species. In the future, it will be used to analyze samples from the Stardust machine, other laboratory analogues and cosmic materials such as meteorites, and IDPs.

## <u>Acknowledgments:</u>

We acknowledge support from the European Research Council under the European Union's Seventh Framework Programme ERC-2013-SyG, Grant Agreement n. 610256 NANOCOSMOS.

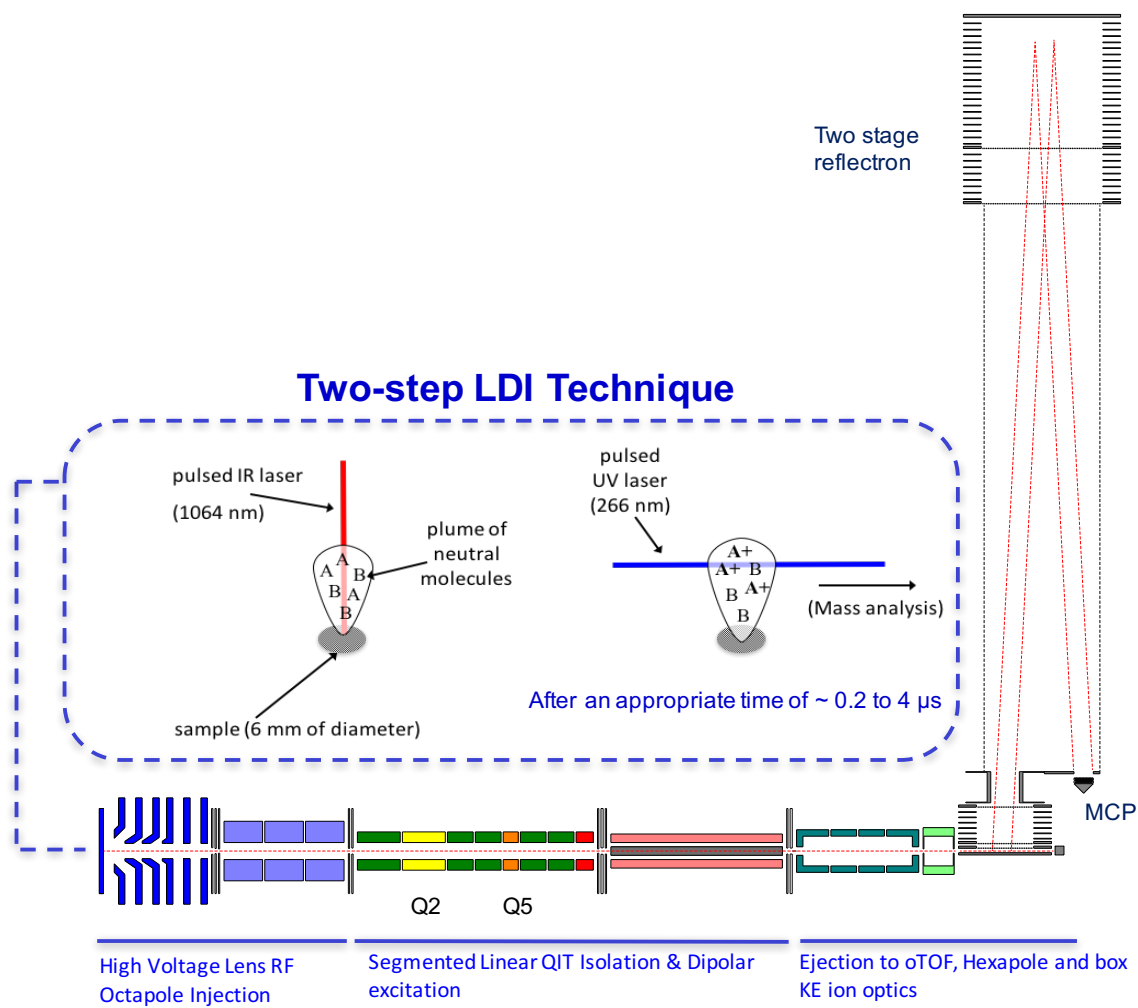

**Figure 1:** Schematic diagram of the AROMA setup highlighting the principal components of the instrument.

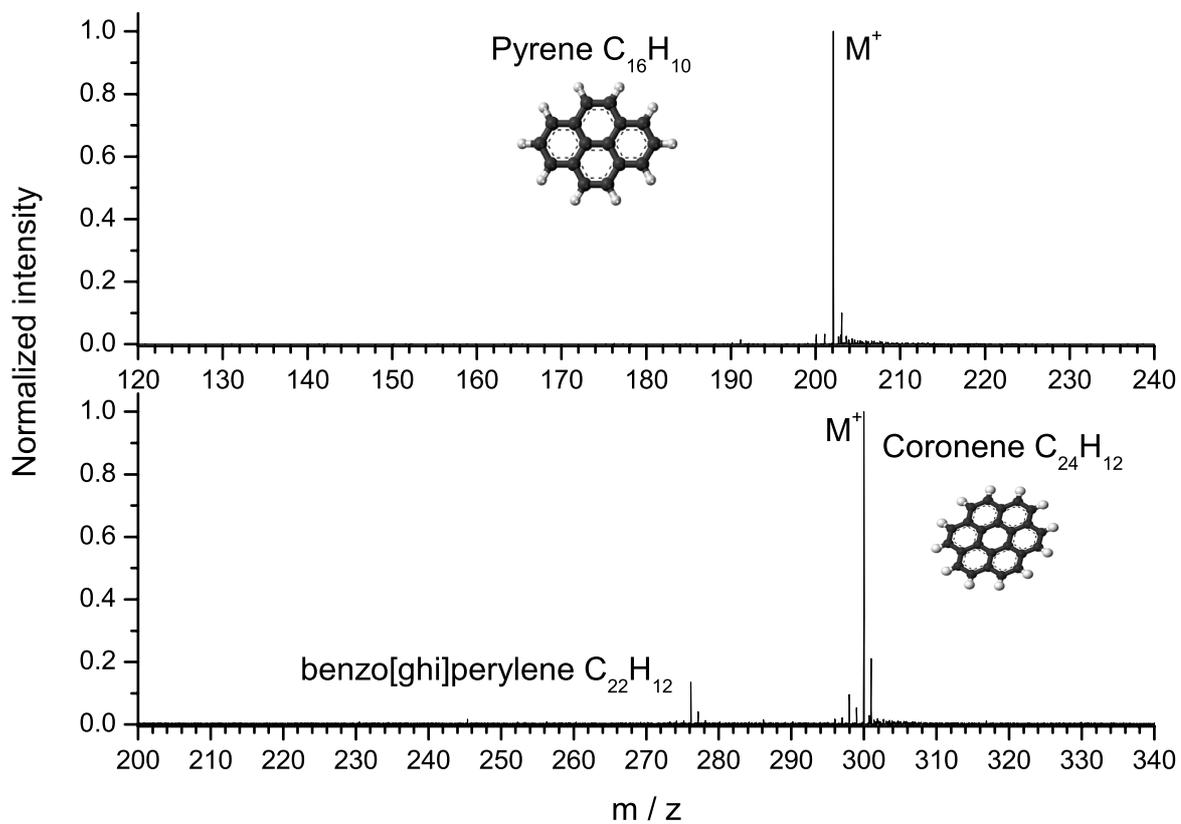

**Figure 2:** AROMA mass spectrum recorded for pyrene (a) and coronene (b). Peak intensities are normalized to the highest peak in the spectrum.

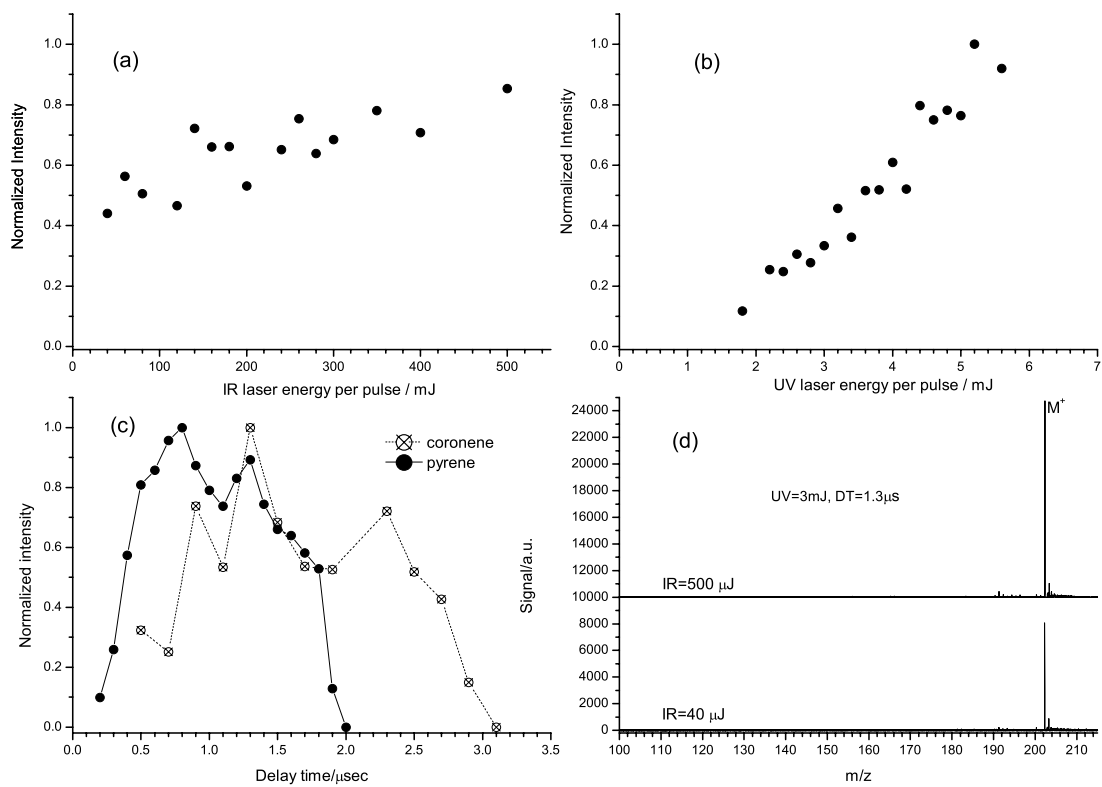

**Figure 3: (a)** Variation of the pyrene parent ion peak intensity as a function of the desorption energy per laser pulse (1064 nm). **(b)** Variation of the pyrene parent ion peak intensity as a function of the ionization energy per laser pulse (266mm). **(c)** Optimization of the delay time between the desorption and the ionization steps. The most appropriate time in this case is 1.3 μs, in order to detect pyrene and coronene cations. In **(a)**, **(b)**, and **(c)** intensities are normalized to the highest value in each graph. **(d)** Two mass spectra recorded for pyrene at the minimum and maximum IR laser pulse energy; No significant change in the spectrum has been observed over the used range.

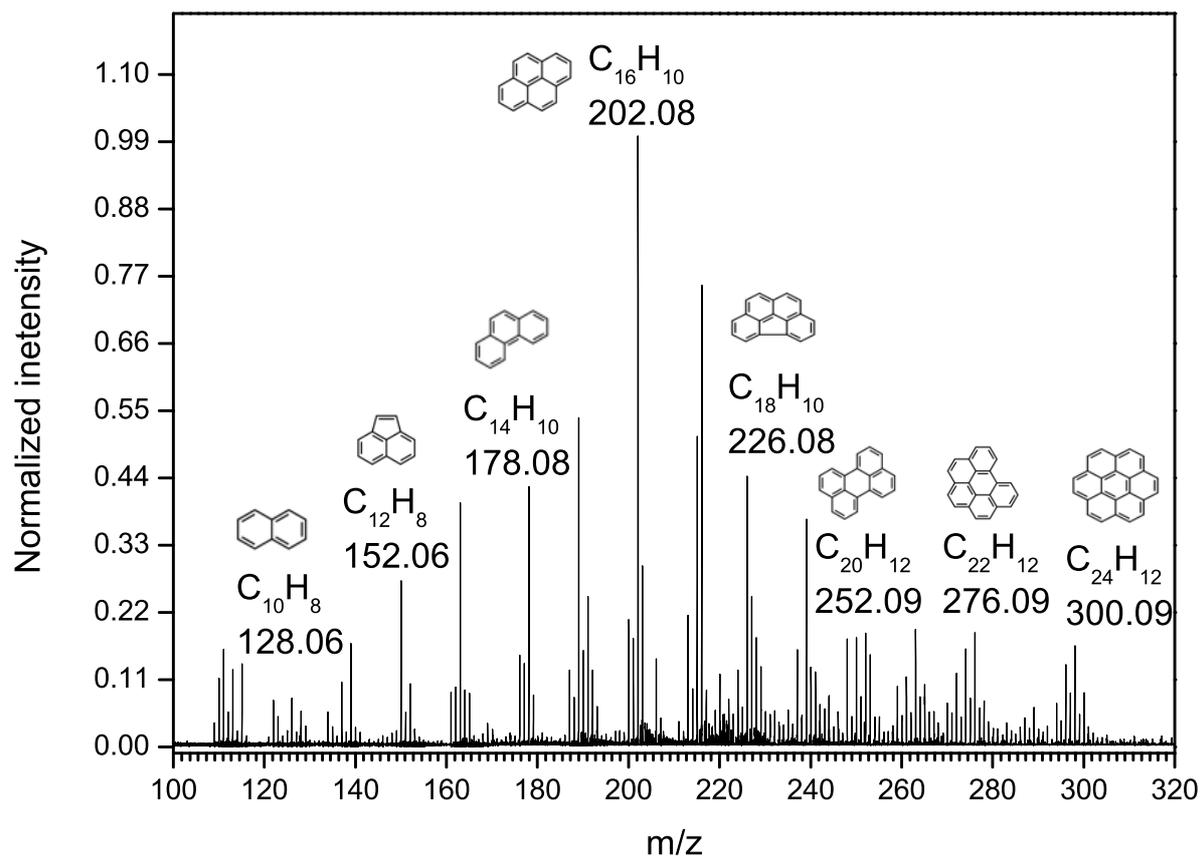

**Figure 4:** PAH distribution recorded using the AROMA setup from one mg of Murchison meteorite powder. Intensities are normalized to the highest peak in the spectrum. Possible isomeric structures are shown for specific peaks.

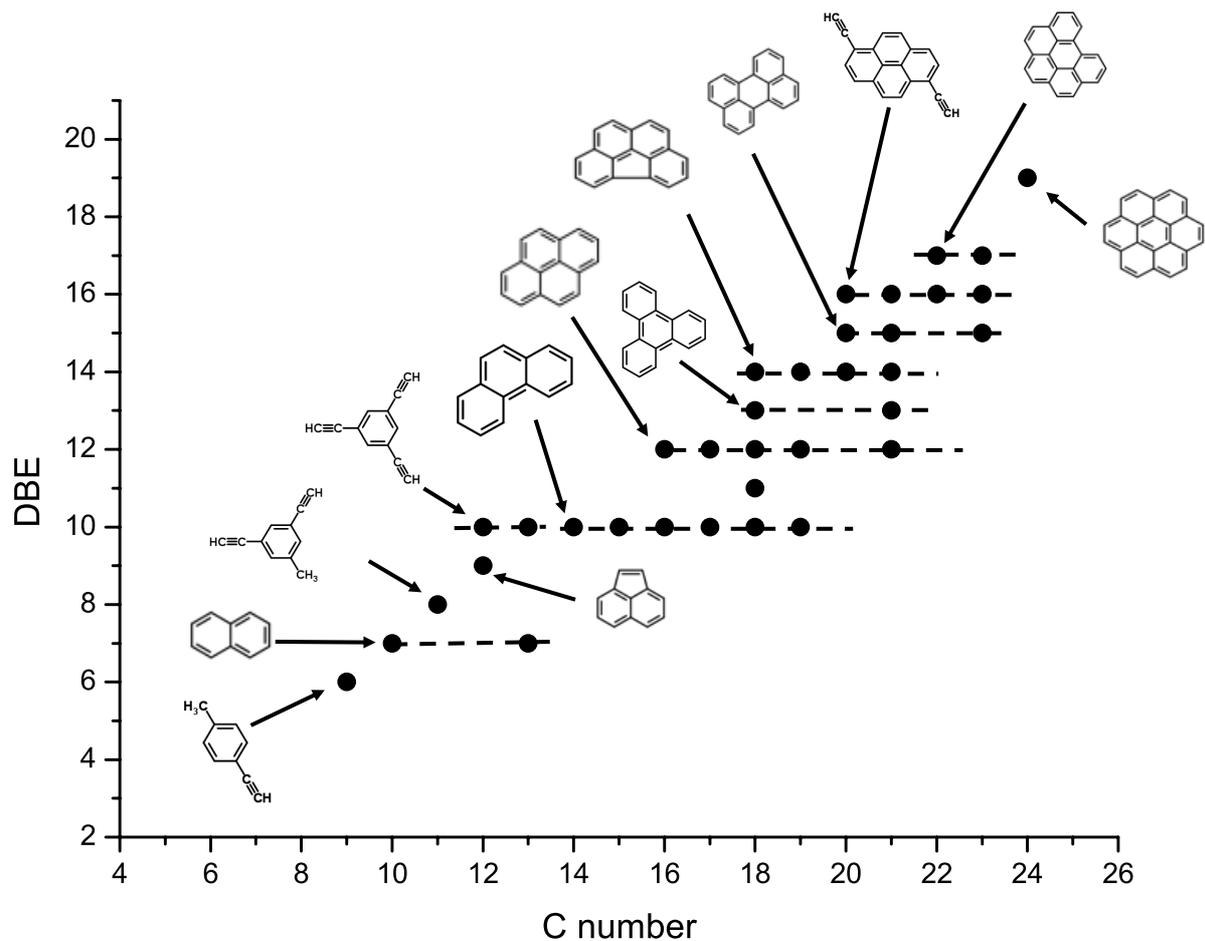

**Figure 5:** DBE versus carbon number plot showing the PAHs detected in the Murchison meteorite and their related products. Species having the same DBE and differing by one carbon number include substitutions by methyl group (-H +CH$_3$). They are shown by dashed lines on the diagram.

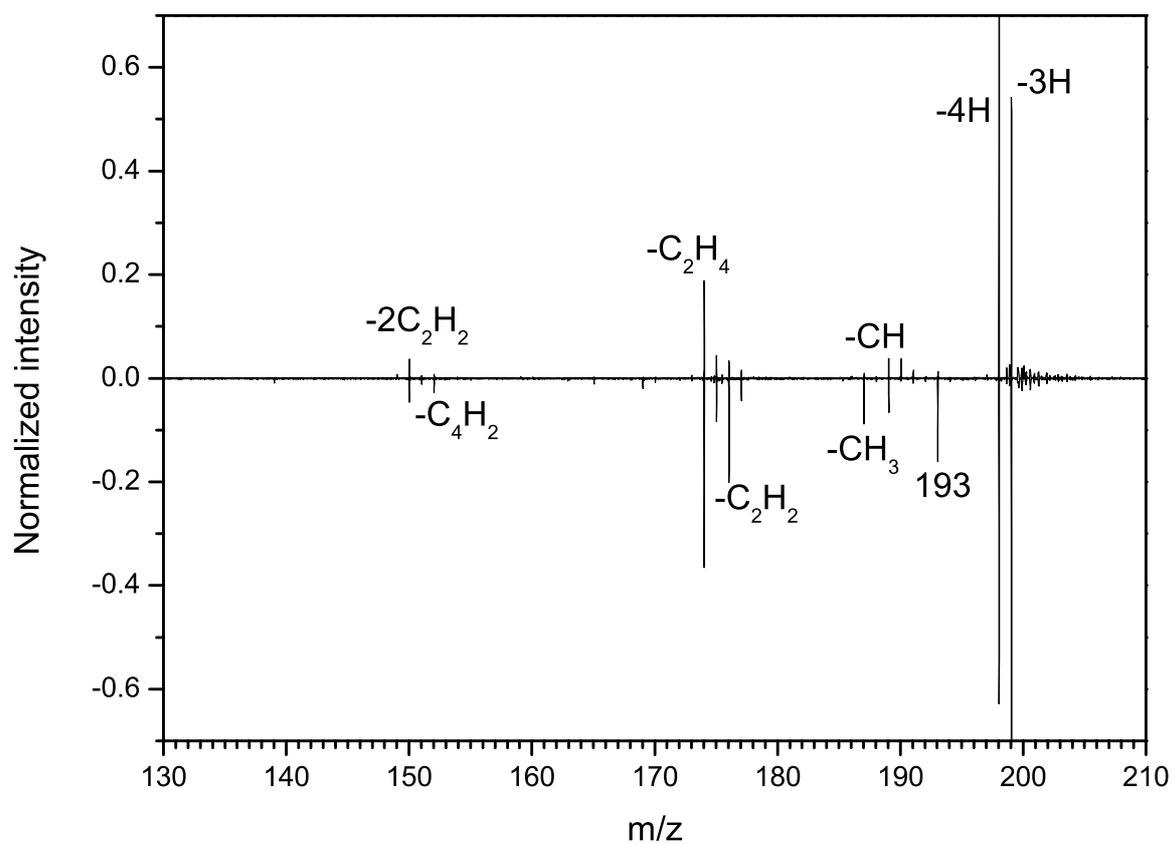

**Figure 6:** CID spectra of m/z = 202 from Murchison meteorite (up spectrum) and pure pyrene sample (down spectrum) recorded with the AROMA setup. Peak intensities are normalized to the highest peak in each the spectrum.

**Table 1 :** Peak assignments for the AROMA mass spectrum of Murchison.

| Mass | Formula | DBE |
|---:|:---:|:---:|
| 116.06 | $C_9H_8$ | 6 |
| 128.06 | $C_{10}H_8$ | 7 |
| 140.06 | $C_{11}H_8$ | 8 |
| 150.04 | $C_{12}H_6$ | 10 |
| 152.06 | $C_{12}H_8$ | 9 |
| 164.06 | $C_{13}H_8$ | 10 |
| 170.11 | $C_{13}H_{14}$ | 7 |
| 178.07 | $C_{14}H_{10}$ | 10 |
| 192.09 | $C_{15}H_{12}$ | 10 |
| 202.07 | $C_{16}H_{10}$ | 12 |
| 206.09 | $C_{16}H_{14}$ | 10 |
| 216.09 | $C_{17}H_{12}$ | 12 |
| 220.12 | $C_{17}H_{16}$ | 10 |
| 226.07 | $C_{18}H_{10}$ | 14 |
| 228.09 | $C_{18}H_{12}$ | 13 |
| 230.11 | $C_{18}H_{14}$ | 12 |
| 232.12 | $C_{18}H_{16}$ | 11 |
| 234.14 | $C_{18}H_{18}$ | 10 |
| 240.09 | $C_{19}H_{12}$ | 14 |
| 246.14 | $C_{19}H_{16}$ | 12 |
| 248.15 | $C_{19}H_{20}$ | 10 |
| 250.07 | $C_{20}H_{10}$ | 16 |
| 252.09 | $C_{20}H_{12}$ | 15 |
| 254.11 | $C_{20}H_{14}$ | 14 |
| 264.09 | $C_{21}H_{12}$ | 16 |
| 266.10 | $C_{21}H_{14}$ | 15 |
| 268.12 | $C_{21}H_{16}$ | 14 |
| 270.14 | $C_{21}H_{18}$ | 13 |
| 272.15 | $C_{21}H_{20}$ | 12 |
| 276.09 | $C_{22}H_{12}$ | 17 |
| 278.11 | $C_{22}H_{14}$ | 16 |
| 290.10 | $C_{23}H_{14}$ | 17 |
| 292.12 | $C_{23}H_{16}$ | 16 |
| 294.14 | $C_{23}H_{18}$ | 15 |
| 300.09 | $C_{24}H_{12}$ | 19 |